# Scanned potential microscopy of edge states in a quantum Hall liquid

Michael T. Woodside[a], Chris Vale[a], Kent L. McCormick[a], Paul L. McEuen[a], C. Kadow[b], K.D. Maranowski[b], A.C. Gossard[b]

[a] *University of California and Lawrence Berkeley National Laboratory, MS 2-200, Berkeley, California, 94720, USA*
[b] *University of California at Santa Barbara, Santa Barbara, California 93106, USA*

Using a low-temperature atomic force microscope as a local voltmeter, we measure the Hall voltage profile in a quantum Hall conductor in the presence of a gate-induced non-equilibrium edge state population at ν=3. We observe sharp voltage drops at the sample edges which are suppressed by re-equilibrating the edge states.

Since the discovery of the quantum Hall effect [1], two-dimensional electron systems (2DES) at low temperatures and high magnetic fields have been found to have a number of unusual properties. One of these is the important role played by the electronic states at the boundary of the 2DES, the edge states of the quantum Hall liquid [2-4]. These quasi-1D extended states form at the edge of the 2DES where the confinement potential causes the energy of filled Landau levels (LLs) to intersect the Fermi level $E_F$. When the effects of electrostatic screening are considered, the picture that emerges is one of compressible strips of partially-filled LLs (variable electron density), occurring where the bulk LLs cross $E_F$ along the 2DES edges, separated by incompressible strips of filled LLs (fixed electron density) [5,6] . Numerous experiments have shown that the properties of the edge states decisively influence transport in a quantum Hall conductor [7,8]. A variety of techniques, including transport measurements [7,8], magnetocapacitance studies [9], photovoltage measurements [10], edge magnetoplasmon studies [11], inductive probes[12], and in-situ single electron transistor electrometers [13], have been used to study the properties of the edge states. Only recently, however, has it been possible to investigate directly the local properties of the quantum Hall liquid, using novel scanned probe techniques [14-16]. We report here measurements of the electrostatic potential at the edge of the quantum Hall liquid made using a cryogenic atomic force microscope.

The system we study is a GaAs/AlGaAs heterostructure grown by molecular beam epitaxy with a 2DES lying 90 nm below the surface [17]. The device is patterned into 10 μm-wide Hall bars by wet chemical etching of the heterostructure. The 2DES has a density of $2.4 \times 10^{15}$ m$^{-2}$ and a mobility of 19 m$^2$/Vs. The device was characterised by standard transport measurements. All measurements were performed at temperatures between 0.7 and 1 K.

Our scanned probe measurements employ a low-temperature atomic force microscope (AFM) [18] operating in non-contact mode which measures the local electrostatic potential in the sample. As shown schematically in Fig. 1, an AC potential $V_0$ at the resonant frequency of the AFM cantilever (30 kHz) is applied to the contacts of the sample, establishing in the sample the AC potential $V(x,y)$ whose distribution is to be measured. This local sample potential $V(x,y)$ interacts electrostatically with the sharp, metallised AFM tip positioned about 50 nm above the sample, causing a deflection of the AFM cantilever. The force on the tip, and hence the deflection of the cantilever, is directly proportional to $V(x,y)$, the local electrostatic potential in the sample. This cantilever deflection is detected with a piezoresistive sensor [19]. A self-resonant positive feedback loop maintains the frequency of the driving voltage $V_0$ at the resonant frequency of the AFM cantilever in order to enhance the force sensitivity, resulting in a sensitivity of about 10 μV/Hz$^{1/2}$. The spatial resolution is approximately 0.2-0.3 μm, limited by the size of the AFM tip (~100 nm diameter) and the distance from the 2DES to the tip. Note that height and contact potential fluctuations [20] cause local variations in the signal strength. We account for these by normalising the measured signal with a simultaneously-measured reference signal whereby a uniform voltage is applied to all contacts on the sample. Further details of the design and operation of the AFM are discussed elsewhere [15,21].

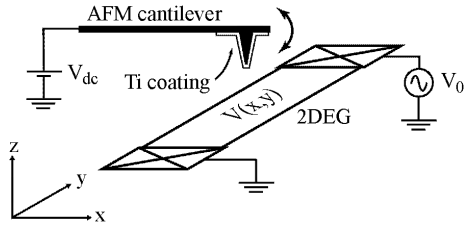

Fig. 1. Schematic diagram of measurement technique. Applying $V_0$ to the sample creates a potential $V(x,y)$ inside the sample which interacts electrostatically with the metallised AFM tip, causing the AFM cantilever to vibrate by an amount proportional to $V(x,y)$.

Previous studies of the local Hall voltage distribution in a quantum Hall conductor underlined the influence of the edge states on the Hall voltage profile [15]. Sharp voltage gradients at the sample edges were observed when transport measurements indicated the presence of edge states out of equilibrium with the bulk. In the current measurements, we focus more closely on the behaviour at the sample edges by controlling the edge state population using gates on the sample. By changing the electron density beneath the gates, we selectively backscatter some of the edge states and establish a non-equilibrium population downstream of the gates (see Fig. 2) [8]. Standard transport measurements (not shown) confirm the presence of a non-equilibrium edge state population, although there is significant inter-edge scattering due to the low sample mobility. In order to maximise the equilibration length, all measurements are taken at a bulk filling factor of $\nu=3$, which is known to support disequilibrated states over long distances [8].

The results of scanning the AFM tip at $\nu=3$ across the Hall bar about 5-10 μm downstream of the gate are shown in Fig. 3 for three different gate voltages. When the gate is open, all of the Hall voltage drops in the bulk of the sample with a slightly non-uniform distribution (Fig. 3a). When the gate voltage is set to backscatter the $\nu=3$ edge state, the potential in the bulk flattens out somewhat and a sharp voltage gradient develops at one edge of the Hall bar, the edge where the backscattered state flows (Fig. 3b). Approximately half of the Hall voltage drop occurs at this edge; the rest occurs in the bulk. When the gate is entirely closed off, the Hall voltage profile is flat (Fig. 3c). The effect of the gate voltage on the Hall voltage profile at the edge of the 2DES can be seen more clearly in an expanded view of the edge (Fig. 3e,f). The voltage gradient arising from the backscattering of the $\nu=3$ edge channel drops over a distance of 0.3 μm, about 0.2 μm from the edge.

These observations can be readily understood in terms of standard theories of a quantum Hall conductor. When the gate is open and all edge states pass through, the edge states are all at the same potential, and there is no voltage drop at the edges (Fig. 3d). Instead, the Hall voltage drops in the bulk of the sample (Fig. 3a), where the Hall voltage distribution is determined by the local conductivity of the states at $E_F$ [15]. When the gate is fully pinched off, all of the edge states are reflected and hence no Hall voltage is observed (Fig 3c). When the gate reflects only the $\nu=3$ edge state, however, the outer edge states downstream of the gate are at potential $V_0$ while the innermost state is at potential 0, as in Fig. 2. This gives rise to a sharp voltage drop across the incompressible strip separating the $\nu=2$ and $\nu=3$ edge states (Fig. 3e). The length over which the voltage drops suggests that the incompressible strip is at most ~300 nm wide. This result agrees well with other measurements [13], but it is close to the resolution limit of the measurement and hence should be viewed as an upper bound on the width of the incompressible strip [22]. Note also that not all the Hall voltage drops across the incompressible strip (Fig 3b), because of inter-edge scattering that occurs between the gate and the measurement location. Transport measurements indicate that the edge states are mostly equilibrated after travelling 40 μm. Hence noticeable equilibration is expected over the 5-10 μm distances in the AFM measurements.

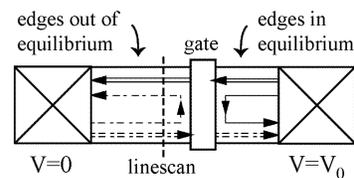

Fig. 2. Measurement configuration. Applying a voltage to the gate reduces the electron density locally, backscattering the innermost edge state (the bulk filling factor shown here is $\nu=3$). The AFM scans across the Hall bar downstream of the gate, where a non-equilibrium edge state population is established.

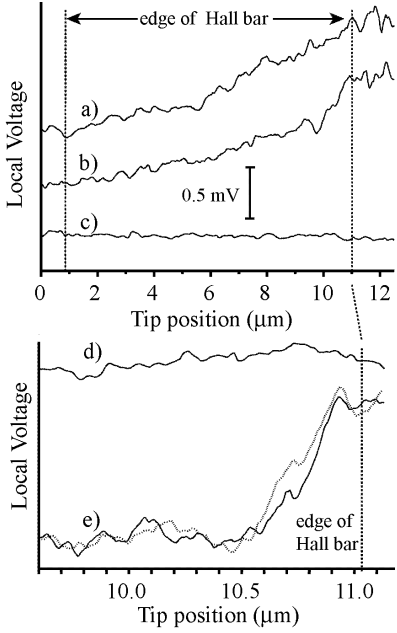

Fig. 3. Hall voltage profiles across 10 μm wide Hall bar at ν=3 (traces offset for clarity). (a) With the gate open, $V_H$ drops in the bulk of the sample. Voltage gradients in the bulk are due to non-uniform local conductivity in the states at $E_F$. (b) With the gate at ν=2, a sharp voltage gradient develops at the edge of the Hall bar where the non-equilibrium edge states flow. Only half of $V_H$ drops at the edge, due to inter-edge scattering. (c) When the gate is pinched off entirely so that all edge states are reflected, the Hall voltage disappears. (d), (e) Close up of the Hall voltage profile at the edge of the sample. There is no gradient at the edge when the gate is open (d). When the gate is at ν=2, the Hall voltage drops over ~ 300 nm, about 200 nm from the edge of the Hall bar. Two traces are plotted to show the reproducibility of the signal (the small-scale features are noise).

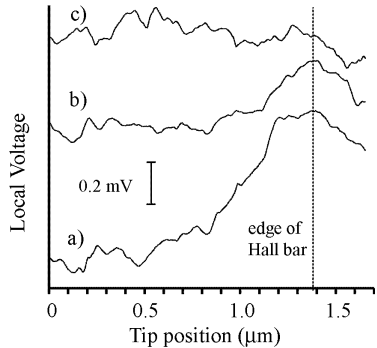

Fig. 4. Effect of DC Hall voltage on voltage profile near edge of Hall bar at ν=3 (traces offset for clarity). (a) The disequilibrated edge states are clearly seen with 0 DC bias. (b) At 5 mV DC bias, little of the voltage gradient is left and the edge states are mostly equilibrated. (c) At 8 mV DC Hall voltage, the edge states are completely equilibrated and no gradient is observed at the sample edge.

Having established edge states out of equilibrium, the edge state population can be re-equilibrated by applying a DC Hall voltage of the order of the LL splitting, $\hbar\omega_c$ [23]. The voltage near the sample edge in the presence of a DC Hall voltage is shown in Fig. 4. All linetraces have the gate reflecting the ν = 3 edge state. As the DC bias is increased from 0 (Fig. 4a) to 5 mV (Fig. 4b), just below $\hbar\omega_c = 5.5$ meV, the voltage drop due to the disequilibrated edge states is reduced substantially, indicating significant re-equilibration. At 8 mV DC bias, well above $\hbar\omega_c$, there is no voltage drop at the sample edge (Fig. 4c), and the edge states are completely equilibrated.

Finally, we studied the perturbation of the sample induced by the measurement technique. The DC voltage applied to the AFM tip in order to detect the local potential in the sample changes the density of the 2DES by ≤ 10% at typical operating conditions [15, 21]. Studies of the effect of the tip voltage, however, show no qualitative change in the measured Hall voltage profile for perturbations of the 2DES ranging from 5% to 20%. In particular, there seems to be little or no equilibration of the edge states induced by the AFM tip, even at tip voltages twice those in the measurements above. We are thus confident that our measurement is not qualitatively affecting the non-equilibrium edge state population we are studying.

In conclusion, we have measured the local Hall voltage across a quantum Hall conductor in the presence of gate-induced disequilibrated edge states at ν = 3. We observe a sharp voltage drop at the edge of the sample along which flow the non-equilibrium edge states. This voltage gradient can be suppressed by equilibrating the edge states, either with the gate or with a DC Hall voltage of order $\hbar\omega_c$. Future work will include more detailed studies of inter-edge channel equilibration and measurements of the local compressibility of the edge of the quantum Hall liquid.

This work was supported by the NSF, NSERC, the AT&T Foundation, and the Packard Foundation.  We acknowledge the Berkeley Microlab for sample fabrication.